# High-throughput fast full-color digital pathology based on Fourier ptychographic microscopy via color transfer

**Yuting Gao,**[a,b] **Jiurun Chen,**[a,b] **Aiye Wang,**[a,b] **An Pan,**[a,*] **Caiwen Ma,**[a,*] **Baoli Yao**[a]

[a] Xi'an Institute of Optics and Precision Mechanics, Chinese Academy of Sciences, Xi'an 710119, China
[b] University of Chinese Academy of Sciences, Beijing 100049, China

**Abstract**. Full-color imaging is significant in digital pathology. Compared with a grayscale image or a pseudo-color image that only contains the contrast information, it can identify and detect the target object better with color texture information. Fourier ptychographic microscopy (FPM) is a high-throughput computational imaging technique that breaks the tradeoff between high resolution (HR) and large field-of-view (FOV), which eliminates the artifacts of scanning and stitching in digital pathology and improves its imaging efficiency. However, the conventional full-color digital pathology based on FPM is still time-consuming due to the repeated experiments with tri-wavelengths. A color transfer FPM approach, termed CFPM was reported. The color texture information of a low resolution (LR) full-color pathologic image is directly transferred to the HR grayscale FPM image captured by only a single wavelength. The color space of FPM based on the standard CIE-XYZ color model and display based on the standard RGB (sRGB) color space were established. Different FPM colorization schemes were analyzed and compared with thirty different biological samples. The average root-mean-square error (RMSE) of the conventional method and CFPM compared with the ground truth is 5.3% and 5.7%, respectively. Therefore, the acquisition time is significantly reduced by 2/3 with the sacrifice of precision of only 0.4%. And CFPM method is also compatible with advanced fast FPM approaches to reduce computation time further.

*An Pan**, E-mail: panan@opt.cn  Caiwen Ma, E-mail: cwma@opt.ac.cn

## 1 Introduction

In the field of biomedicine, accurate and efficient observation of pathologic slices is of great significance to cell morphology detection, pathologic analysis and disease diagnosis, which acts as the bridge between fundamental research and clinical applications [1]. It is considered the gold standard for tissue-based diagnostics, with some well-established versions of common stains, such as Masson [2], rosein, Sirius red [3], hematoxylin and eosin (H&E) [4] and oil red [5], having been used for over a hundred years. Therefore, on the one hand, pathologic slices are usually stained for specific recognition, given the fact that humans are more sensitive to color information and have established the habit of classification according to its color as shown in Fig. 1. On the other hand, digital pathology that uses digital cameras to collect stained pathologic slices improves the imaging efficiency compared with naked eyes and reduces the overlook and double counting. However, there is a tradeoff between high resolution (HR) and wide field-of-view (FOV) in digital pathology, resulting in the artifacts of scanning and stitching.

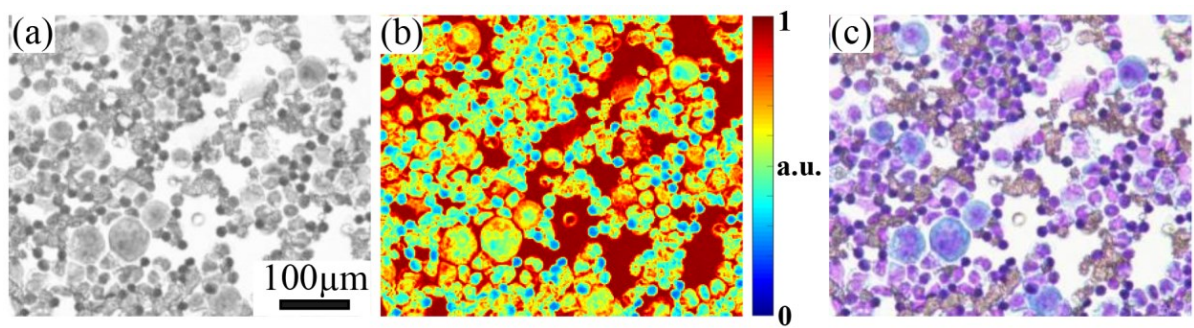

**Fig. 1** The importance of true colorization of grayscale images. (a) Grayscale image of a cell section. (b) Pseudo-color image (c) True-color image of the cell section.

Fourier ptychographic microscopy (FPM) invented in 2013 by Zheng and Yang et al. [6-8] is a promising computational imaging technique that eliminates these artifacts in digital pathology and provides higher throughput, sharing its root with phase retrieval [9-11] and aperture synthesis [12]. The insight is that the objective can only collect light ranging from a certain angle, characterized by the numerical aperture (NA). However, parts of the scattering light with higher angle illumination can be also collected due to light-matter interaction and the sample's high-frequency information can be modulated into the passband of the objective lens. Instead of conventionally stitching small HR image tiles into a large FOV, FPM uses a low NA objective to take advantage of its innate large FOV and stitches together low resolution (LR) images in Fourier space to recover HR. Due to its flexible setup, performance without mechanical scanning and interferometric measurements, FPM has found successful applications in the digital pathology and whole slide imaging systems [13-15]. And FPM is developed toward high-throughput imaging [16-18], high-speed and single-shot imaging [19-21], three-dimensional (3D) imaging [22-24], mixed-state decoupling [25, 26], biomedical applications [27, 28], optical cryptosystem [29], and remote sensing [30].

In order to achieve high-throughput full-color digital pathology based on FPM, generally, there are four kinds of methods. The first one is to restore a HR image with a monochrome camera at tri-wavelengths respectively and synthesize them into one HR full-color image. It is easy to achieve with a programmable R/G/B LED array and there are no extra limitations on the overlapping ratio and system environments compared with the original FPM. But the drawbacks are also obvious that it is time-consuming and needs to calibrate the intensity of different wavelengths, i.e., white balance. The second method is to use a chromatic camera to separate three primary channels with Bayer filter, which is timesaving and does not have extra limitations on the overlap ratio and system environments either. But the pixel size of a chromatic camera is usually larger than the monochrome camera, otherwise, the photon efficiency will be fairly low. Also, it needs to correct the error of color leakage [31]. The third approach is termed wavelength-multiplexed FPM (WMFPM) scheme with a monochrome camera and multi-wavelength simultaneous illuminations [25, 31-34]. Seemingly, it keeps using a monochrome camera and reduces the acquisition time by 2/3. But around 3 times higher overlapping ratio is required, therefore Sun et al. found that it cannot reduce so much time [33]. And it adds the algorithm complexity for decoupling, which makes it hard to calibrate the system errors together. Much worse, three extra LR images of tri-wavelengths are required respectively to avoid the convergence to roughly the same gray value [33, 34]. And the final approach is deep learning based on convolutional neural network (CNN), pixel-to-pixel transformation approach or generative adversarial network (GAN) [35, 36]. However, there are several inevitable problems, such as overfitting, difficulty in generalization and obtaining abundant training sets, lack of physical principles and so on [35-38].

Herein, inspired by the concept of color matching [39], a color transfer FPM, termed CFPM was reported. The insight is to change the original RGB three channels into two channels, a grayscale channel and a color texture channel. The color texture information of a LR full-color pathologic image is directly transferred to the HR grayscale FPM image captured by only a single wavelength, which may be regarded as a kind of unsupervised transfer learning. Note that there is an optimal selection for illumination wavelength to obtain the highest contrast for the grayscale FPM image, which depends on the absorption of dyes and is easy to access as a priori knowledge. Different FPM colorization schemes were analyzed and compared with thirty different biological samples. The average root-mean-square error (RMSE) of conventional method and CFPM compared with the ground truth is 5.3% and 5.7%, respectively. Therefore, the acquisition time is significantly reduced by 2/3 with the sacrifice of precision of only 0.4%, which achieves a leap in imaging speed while ensuring the quality of image recovery. And CFPM method is also compatible with advanced fast FPM approaches [18, 33] to reduce computation time further. In addition, the color space of FPM based on the standard CIE-XYZ color model and display based on the standard RGB (sRGB) color space were established and illustrated in detail. And other potential parallel schemes are also discussed and compared.

## 2    Principle of color transfer FPM (CFPM)

Color matching is to apply the hue of the donor image to the acceptor image so that the acceptor image has the same hue as the donor image [39], as shown in Fig. 2(a) as an example. It allows the users to make their style of paintings the same as the style of famous painters whomever they want. Inspired by this concept, a color transfer scheme was reported by transferring the color texture information of a LR full-color image to a HR FPM reconstructed image, as shown in Fig. 2(b).

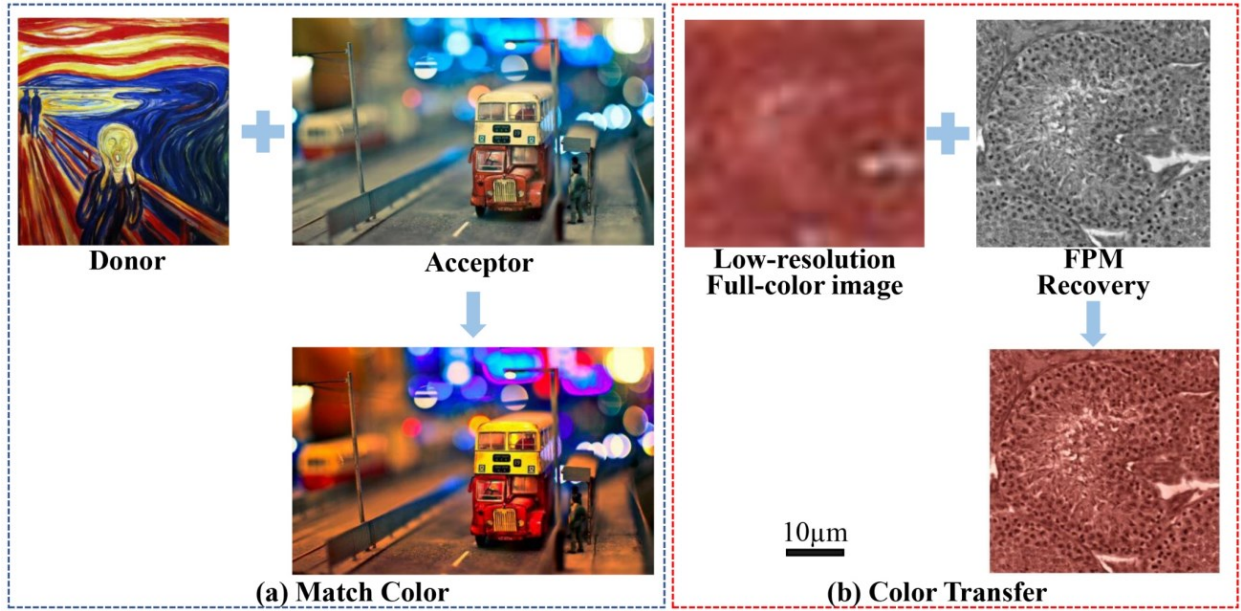

**Fig. 2** Inspiration of color matching and the principle of CFPM approach. (a) Match color: transfer the hue of donor image to the acceptor image. (b) Color transfer: transfer the color texture of donor image to the acceptor image.

Specifically, the color transfer scheme contains three steps as shown in Fig. 3. First, the *sRGB* color space of donor image is converted to *Lab* color space, which was invented by Ruderman et al. in 1988 [40], where *L* is the channel representing brightness with values ranging from 0 to 100 as measurement, *a* represents the red and green channels and *b* represents the yellow and blue channels, both of whose color ranges are [+127, -128]. Compared with the *sRGB* color space, the *Lab* color space separates the brightness and color texture into two channels completely, so that they will not interfere with each other and is much convenient for color transfer. Second, the color texture information of the donor is transferred to the acceptor image within the *Lab* color model by matching the brightness histogram of the acceptor image to the donor image to eliminate the overall difference between them. Third, the *Lab* color model of CFPM recovery is converted to *sRGB* color space for display.

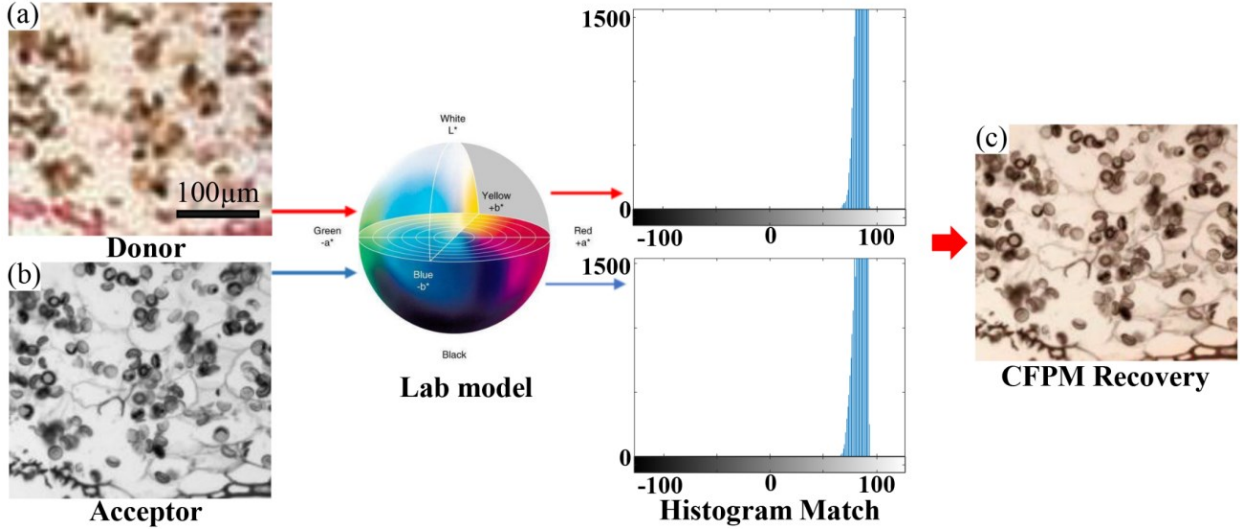

**Fig. 3** Schematic diagram of CFPM. (a) Donor image: A LR full-color image with the same FOV of FPM captured by the same objective. (b) Acceptor: FPM recovery with a single wavelength. (c) CFPM final recovery.

In order to convert the *sRGB* color model to the *Lab* color model, the detailed steps are as follows.

**Step 1** Convert *sRGB* color space to *LMS* color space represented by the responses of three cones of human eyes, which is given by [40]:

$$\begin{pmatrix} L \\ M \\ S \end{pmatrix} = \begin{pmatrix} 0.3811 & 0.5783 & 0.0402 \\ 0.1967 & 0.7244 & 0.0782 \\ 0.0241 & 0.1228 & 0.8444 \end{pmatrix} \begin{pmatrix} R \\ G \\ B \end{pmatrix} \quad (1)$$

where $L$ is the long-wavelength channel, $M$ is the medium wavelength channel and $S$ is the short wavelength channel.

**Step 2** In order to make *LMS* color space data more concentrated and more in line with the human perception of color, the *LMS* color space is converted to its logarithmic space with base 10.

$$\begin{cases} L' = \log L \\ M' = \log M \\ S' = \log S \end{cases} \quad (2)$$

**Step 3** Use the principal component analysis (PCA) of the data to obtain the conversion parameters from the *LMS* color model to the Lab color model as follows [40].

$$\begin{pmatrix} L \\ a \\ b \end{pmatrix} = \begin{pmatrix} 1/\sqrt{3} & 0 & 0 \\ 0 & 1/\sqrt{6} & 0 \\ 0 & 0 & 1/\sqrt{2} \end{pmatrix} \begin{pmatrix} 1 & 1 & 1 \\ 1 & 1 & -2 \\ 1 & -1 & 0 \end{pmatrix} \begin{pmatrix} L' \\ M' \\ S' \end{pmatrix} \quad (3)$$

And the inverse transformation from *Lab* color space to *sRGB* color space is the same principle as above. After the preprocessing of donor and acceptor images, the color transfer is detailed as follows. Note that both the FOV of the donor image and acceptor image are the same, while the resolution is not the same, which has a difference in upsampling magnification.

**Step 1** Calculate the brightness value *PD* of each pixel of the donor image and the brightness variance value *DD* in the 5×5 neighborhood and define the parameter as *RD*=0.5*PD*+0.5*DD*.

**Step 2** For a pixel *j* of the acceptor image, calculate the brightness value $PA_j$ and the brightness variance value $DA_j$ in the 5×5 neighborhood to obtain $RA_j$, where $RA_j$=0.5$PA_j$+0.5$DA_j$.

**Step 3** Subtract $RA_j$ of the acceptor image by the *RD* value of each pixel of the donor image and find the pixel of the donor image at the minimum, which is the point closest to the *j-th* pixel of the acceptor image. Then assign the *a* and *b* channel values of this pixel of donor image to the pixel *j* of the acceptor image.

**Step 4** Repeat the same operations for each pixel of the acceptor image, and we can obtain the color transfer results.

Tested by our experiments, the 5×5 neighborhood is the best choice to resist the noise and ringing effect of coherent images. There are several typical solutions that are based on the Welsh algorithm [41, 42] to realize the colorization of grayscale images. Different from them, on the one hand, FPM is a coherent imaging technique while the donor image of the Welsh algorithm is captured by incoherent imaging, therefore, there is a wavelength selectivity in FPM. On the other hand, the application scenarios are totally different. The kind of Welsh algorithm is applied in macroscopic scale and the donor image is a similar reference picture, while FPM is applied in digital pathology which is microscopic scale, and the donor image is the object itself with the same FOV.

## 4 Experiments

*4.1 Establishment of color space and display*

The FPM procedure can be referred as Ref. 6-8 and will not be detailed here. For a finite set of illumination vectors $k_{m,n}$, each intensity image is denoted by:

$$I_i(r) = \left| \mathcal{F}_{[O(k-k_i)P(k)]}(r) \right|^2 \tag{4}$$

where $r=(x, y)$ denotes the lateral coordinates in the sample plane, $k=(k_x, k_y)$ denotes the lateral coordinates in the Fourier domain, $k_i=(k_{x,i}, k_{y,i})$ is the spatial frequency of the local plane wave emitted by each LED; $P(k)$ is the pupil function, $O(k-k_i)$ represents the exit wave at the pupil plane, and subscript $i$ is the sequence number of LR images; $\mathcal{F}\{\cdot\}$ is the 2D Fourier transform. The schematic and experimental configurations of FPM are shown in Figs. 4(a) and 4(b), respectively. A 32×32 programmable R/G/B LED array (Adafruit, 4 *mm* spacing, controlled by an Arduino) is placed at 70.0 *mm* above the sample (Fig. 4(a1)), while 15×15 center LED elements are lighted up sequentially for the data acquisition process. Tested by the spectrometer of Ocean Optics, the red, green, and blue LEDs have a dominant narrow peak at the wavelength of 630.1 *nm*, 515.0 *nm*, and 462.6 *nm* within the full width at half maximum (FWHM) of 20.8 *nm*, 38.0 *nm*, 34.6 *nm*, respectively (Figs. 4(c1)-3(c4)). A compact inverted microscope is used as shown in Fig. 4(a2) with a light path diagram, which can be further combined with fluorescence imaging easily (Fig. 4(a3)) [43]. All the data are captured by a 4×/0.1NA plan achromatic objective and a 16-bits sCMOS camera (Neo 5.5, Andor, 2160×2560 pixels, 6.5 *μm* pixel pitch). To achieve the FPM reconstruction, the noise, system error and vignetting should be considered carefully [44-46].

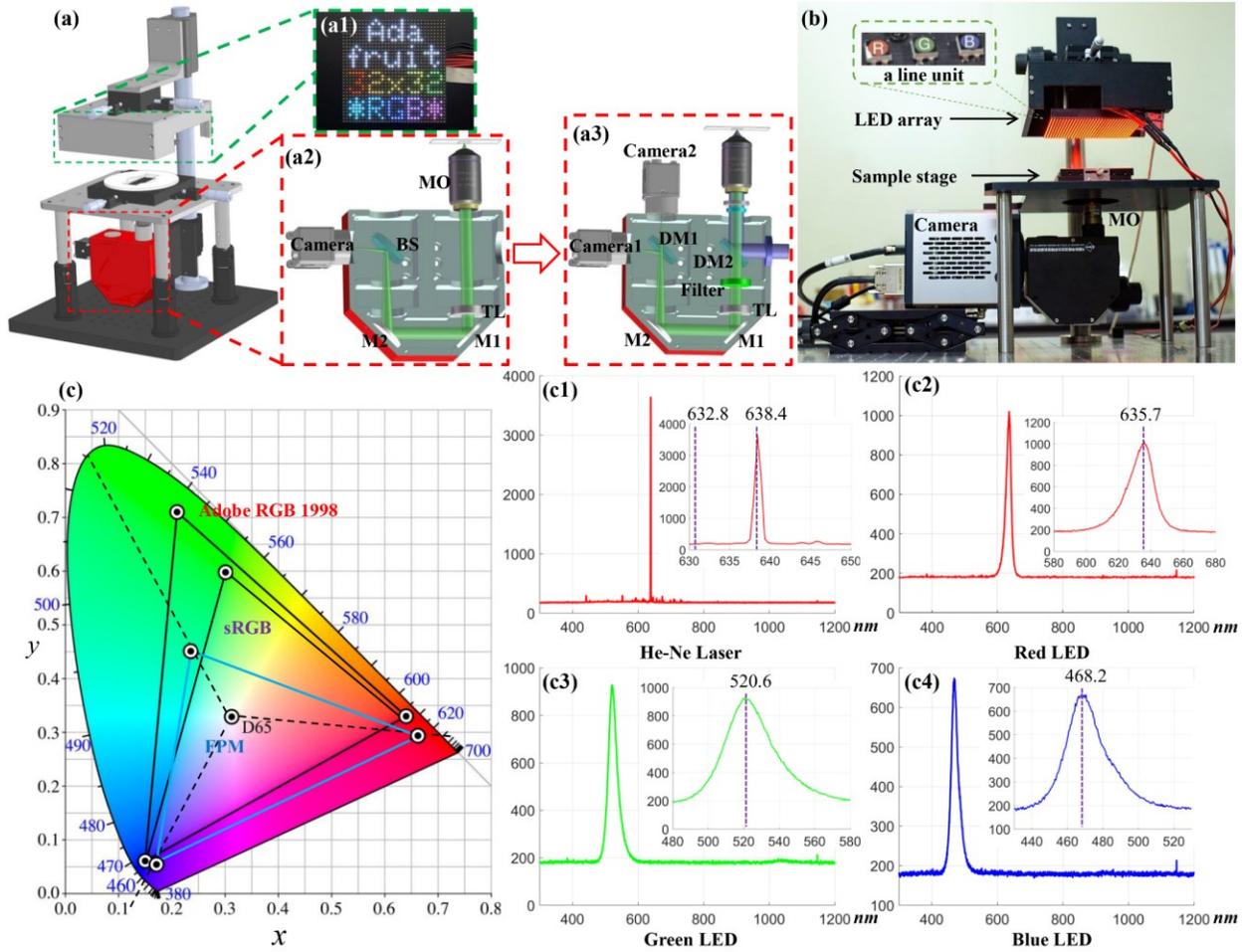

**Fig. 4** CFPM setup with the illumination of LED board. (a) Schematic and its experimental photograph (b). (a1) A 32×32 programmable R/G/B LED matrix. (a2) The enlargement of a compact inverted microscope with light path diagram. It can be simply modified to realize multimodal imaging (a3). MO: microscope objective; TL: tube lens; M1 and M2: mirrors; BS: beam splitter; DM1 and DM2: dichroic mirror. (c) Standard color gamut chart of 1931 CIE-XYZ and our color space of FPM (blue triangle). (c1-c4) Wavelength and intensity calibration. He-Ne laser: 5.6 *nm* deviation tested by the spectrometer of Ocean Optics. Red LED: 630.1 *nm* of center wavelength, 20.8 *nm* of FWHM.

Green LED: 515.0 *nm* of center wavelength, 38.0 *nm* of FWHM. Blue LED: 462.6 *nm* of center wavelength, 34.6 *nm* of FWHM.

In order to obtain the correct and accurate true-color images, the establishment of color space, white balance and the transformation for display of different color spaces are significant. Correspondingly, the three LED light sources are calibrated on the 1931CIE-XYZ standard chart [47] as three primary colors (Fig. 4(c)). Then the color space of our FPM system can be obtained, as shown in the blue triangle area. The quantitative calculation to obtain the FPM color space is as follows [47]:

$$\begin{aligned} R &= \int_0^\infty I_r(\lambda) r(\lambda) d\lambda \\ G &= \int_0^\infty I_g(\lambda) g(\lambda) d\lambda \\ B &= \int_0^\infty I_b(\lambda) b(\lambda) d\lambda \end{aligned} \quad (5)$$

where $I_r(\lambda)$, $I_g(\lambda)$, $I_b(\lambda)$ are the spectrum curves of three primary colors to be calibrated, namely the spectrum curves in Figs. 4(c1)-(c4), and $r(\lambda)$, $g(\lambda)$, $b(\lambda)$ are the normalized matching functions of CIE-RGB color space. Note that the color matching function will have negative values, which is not easy to operate, the CIE-RGB color space is therefore usually converted to CIE-XYZ color space with the conversion matrix defined as [47]:

$$\begin{bmatrix} X \\ Y \\ Z \end{bmatrix} = \begin{bmatrix} 2.7688 & 1.7517 & 1.1301 \\ 1.0000 & 4.5906 & 0.0601 \\ 0 & 0.0565 & 5.5942 \end{bmatrix} \begin{bmatrix} R \\ G \\ B \end{bmatrix} \quad (6)$$

where CIE-XYZ space is still a three-dimensional space, which is not conducive to display and calculation, so the XYZ three-dimensional space is projected onto the $X+Y+Z=1$ plane (Fig. 4(c)). Therefore, our FPM-XYZ coordinate values are (0.6625, 0.2901, 0.0474)$_R$, (0.2410, 0.4521, 0.3069)$_G$, (0.1719, 0.0608, 0.7663)$_B$, respectively. And D65 is a standard white light source named

after its color temperature of 6500 K, which serves as an artificial light source for sunlight simulation. Usually, D65 is regarded as the white balance point, and it's RGB space coordinate value is (0.95047,1.00000,1.08883). The white balance coefficient of each channel can be given by:

$$W = \gamma_1 R + \gamma_2 G + \gamma_3 B \tag{7}$$

where ($\gamma_1$, $\gamma_2$, $\gamma_3$) is the white balance coefficient. Therefore, the white balance coefficient in our FPM color space is:

$$\begin{bmatrix} \gamma_1 \\ \gamma_2 \\ \gamma_3 \end{bmatrix} = \begin{bmatrix} x_R & x_G & x_B \\ y_R & y_G & y_B \\ z_R & z_G & z_B \end{bmatrix}^{-1} \begin{bmatrix} R_W \\ G_W \\ B_W \end{bmatrix} = \begin{bmatrix} 0.6625 & 0.2410 & 0.1719 \\ 0.2901 & 0.4521 & 0.0608 \\ 0.0474 & 0.3069 & 0.7663 \end{bmatrix}^{-1} \begin{bmatrix} 0.95047 \\ 1.00000 \\ 1.08883 \end{bmatrix} = \begin{bmatrix} 0.6308 \\ 1.7136 \\ 0.6956 \end{bmatrix} \tag{8}$$

So far, we have obtained our FPM color space and its white balance coefficient for display, however, monitors or projectors usually use *sRGB* color space, so for a certain color in the FPM-XYZ color space, its corresponding *sRGB* color space coordinate values are:

$$\begin{aligned} C_{FPM-XYZ} &= \begin{bmatrix} x_R & x_G & x_B \\ y_R & y_G & y_B \\ z_R & z_G & z_B \end{bmatrix} \begin{bmatrix} \gamma_1' & 0 & 0 \\ 0 & \gamma_2' & 0 \\ 0 & 0 & \gamma_3' \end{bmatrix} C_{sRGB} = \\ &\begin{bmatrix} 0.64 & 0.30 & 0.15 \\ 0.33 & 0.60 & 0.06 \\ 0.03 & 0.10 & 0.79 \end{bmatrix} \begin{bmatrix} 0.6445 & 0 & 0 \\ 0 & 0.1919 & 0 \\ 0 & 0 & 1.2029 \end{bmatrix} C_{sRGB} = TC_{sRGB} \end{aligned} \tag{9}$$

Where ($\gamma_1'$, $\gamma_2'$, $\gamma_3'$) is the white balance coefficient of *sRGB* color space, ($x_R$, $y_R$, $z_R$) and similar are the coordinate values of *sRGB* color space in the CIE-XYZ map. Then the matrix converting XYZ to *sRGB* is:

$$T^{-1} = \begin{bmatrix} 3.2405 & -1.5371 & -0.4985 \\ -0.9693 & 1.8760 & 0.0416 \\ 0.0556 & 0.2040 & 1.0572 \end{bmatrix} \tag{10}$$

Note that there are negative elements in the matrix, that is, the conversion from FPM-XYZ space to *sRGB* space possibly make a certain component a negative value. It means that all colors in *sRGB* color space should locate inside its triangle, and if a certain color falls outside the triangle, it cannot be displayed by a monitor or projector. Usually, a simple solution is to force values outside the range to fall on the boundary of the range, force negative values to equal 0, and force values greater than 1 to equal 1. This simple approach is equivalent to directly filling in the color on the boundary of the triangle, but it will lose some accuracy of colors. As shown in the color gamut diagram of Fig. 4(c), the final color can be displayed is the overlapping area of FPM color space and sRGB color space. The transformation for display of other color space, e.g., the NTSC color space, is in the same way.

*4.2 CFPM results*

Figures 5 and 6 show two typical results with different dyes. The ground truth is captured by a 10×/0.3NA plan achromatic objective with all the LEDs lighting and synthesized by conventional three channels. To evaluate the performances, the root-mean-square error (RMSE) is used and given by:

$$RMSE = \sqrt{\sum_{x=1}^{X}\sum_{y=1}^{Y}\left(f(x,y)-g(x,y)\right)^2 / (X \times Y)} \qquad (11)$$

where $f(x, y)$ and $g(x, y)$ are two virtual images, $X$ and $Y$ donate the size of two-dimensional images. The LR full-color donor images are synthesized from three LR images with normal incidence and are shown in Fig. 5(a) and Fig. 6(a), respectively. The HR acceptor images are reconstructed by the FPM procedure and are shown in Fig. 5(b) and Fig. 6(b) under a single wavelength, respectively. And the CFPM reconstructions are shown in Figs. 5(d1-d3) and Figs. 6(d1-d3), respectively. Among them, the best performance for stained resting sporangia is under the green channel with the error of 5.37% (Fig. 5), while it is red channel for stained lily bud cells with the error of 4.83% (Fig. 6). Therefore, there is a wavelength selectivity for the CFPM approach, which is relevant to the absorption of dyes. So it is crucial to know the absorption of specifical dye and it is easy to access as a priori knowledge in fact.

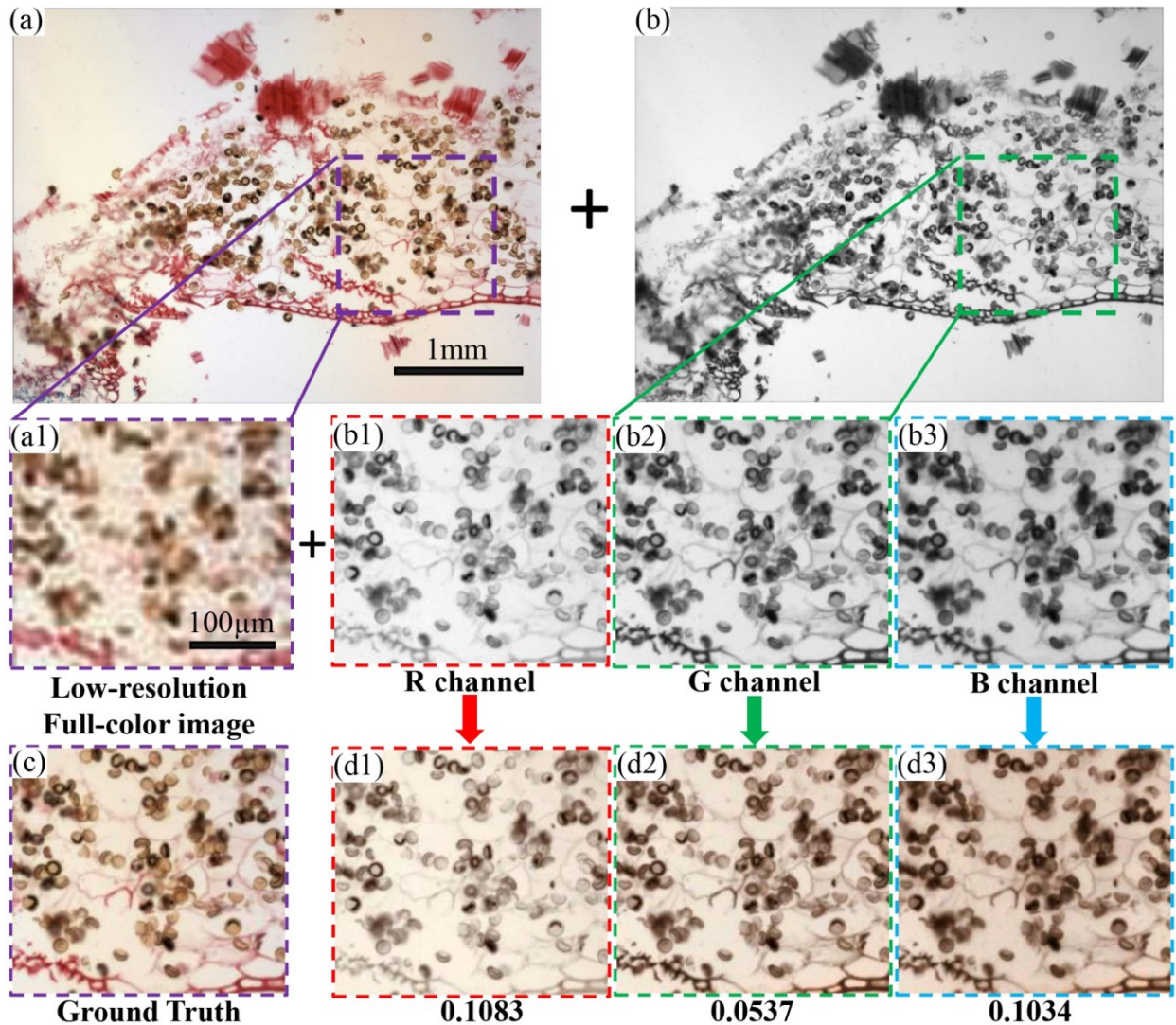

**Fig. 5** Stained resting sporangia. (a) The LR donor image with entire FOV. (b) The FPM recovery image under green channel. (a1) A tile of (a). (b1-b3) Close-ups of FPM recovery images under red, green, blue channel, respectively. (c) Ground truth. (d1-d3) The corresponding results of CFPM respectively.

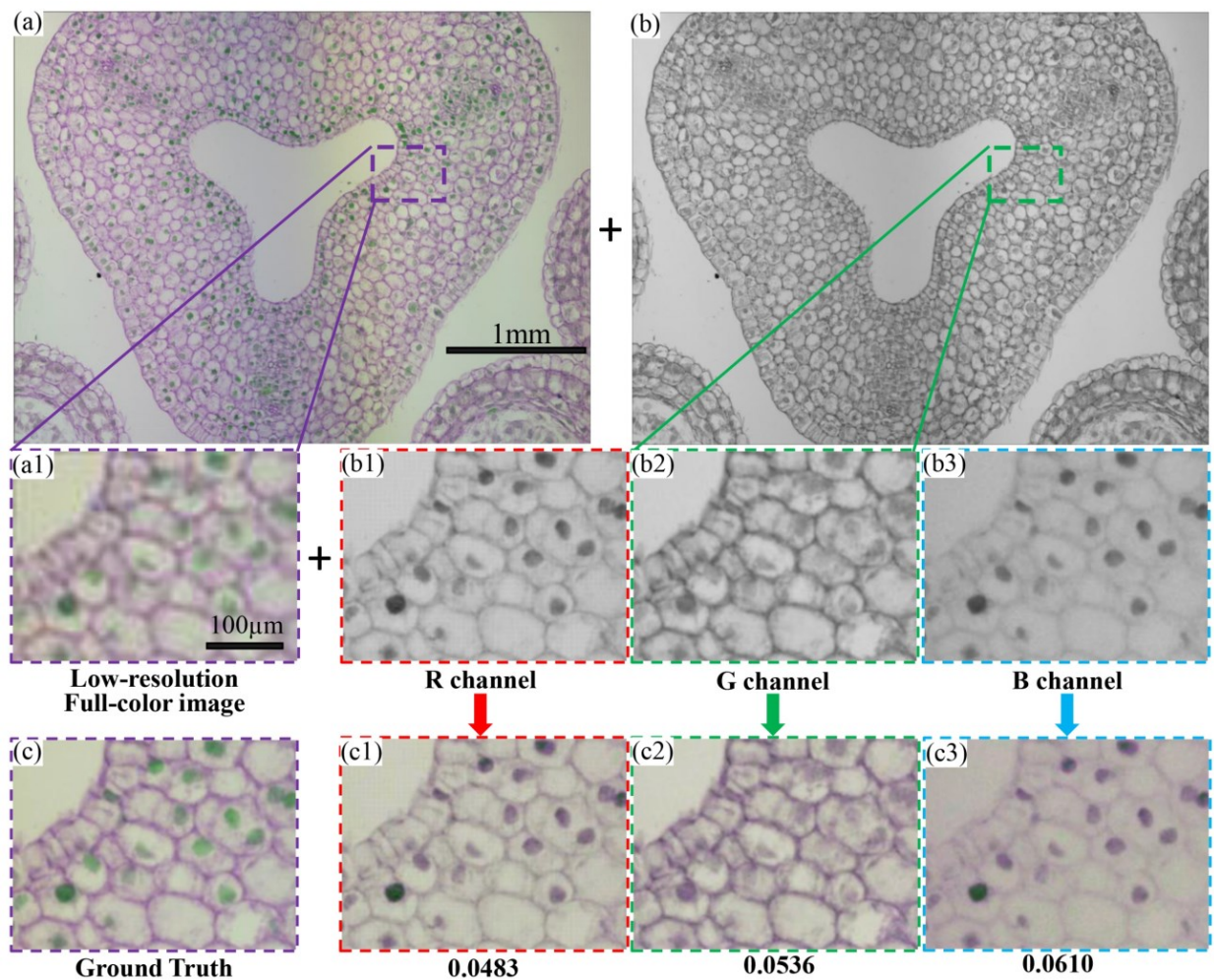

**Fig. 6** Stained lily bud cells. (a) The LR donor image with entire FOV. (b) The FPM recovery image under green channel. (a1) A tile of (a). (b1-b3) Close-ups of FPM recovery images under red, green, blue channel, respectively. (c) Ground truth. (d1-d3) The corresponding results of CFPM respectively.

Figure 7(a) shows the statistical results of 30 different pathologic or biological samples to evaluate the performance of CFPM precisely and different colorization methods are compared. Note that three extra LR images have been added into the WMFPM algorithm. In terms of the average value of RMSE, the difference between the results of conventional R+G +B and CFPM is less than 1%, while that of WMFPM is worse than that of both methods with an average level of 11.85%, which is roughly twice that of RMSE of the former two schemes. Note that sometimes the CFPM would be better than the conventional method. In terms of recovery time that includes the acquisition time and computation time, since FPM reconstruction is only performed on a single channel, the acquisition time of CFPM is around 1/3 of that of conventional R+G+B and the computation time is also 1/3. However, the acquisition time of the WMFPM approach nearly the same as the conventional method or a bit fast due to shorter exposure time, and the computation time is much heavier than the conventional method.

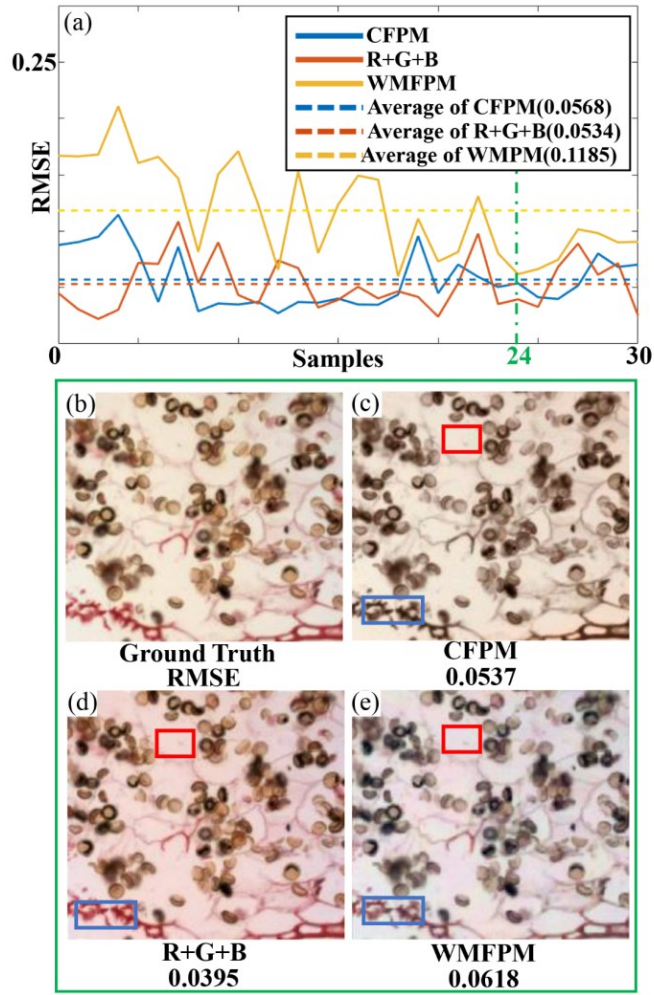

**Fig. 7** Comparisons between three kinds of FPM colorization results. (a) RMSE curves of conventional R+G+B, our CFPM, and WMFPM with 30 pathologic or biological samples. (b) Ground truth of the 24th biological sample. (c-e) Recovery of corresponding CFPM, conventional R+G+B, and WMFPM respectively.

Among the 30 samples, a typical sample whose RMSE result is similar to the average of CFPM is selected for display as shown in Figs. 7(b-d). It should be noticed that even the conventional method will slightly have the chromatic aberration as shown in the region of the red box, while it is better with the CFPM approach for this target. But if two dyes are used together for the specific recognition of different parts, the CFPM may not be able to obtain the right color as shown in the region of the blue box, because there may be totally different absorption for different dyes. Therefore, currently, the CFPM method is better for stained samples with a single dye.

## 5  Discussions

Inspired by the CFPM approach, there may be other color transfer schemes, e.g., transfer a HR full-color image with small FOV to the entire FOV as shown in Fig. 8. The donor image is captured by a 10×/0.3NA plan achromatic objective with bright-field microscope to ensure to obtain a clear tile. Figures 8(b-d) are the other three tiles of HR FPM reconstruction under the best green channel. Using the donor image (Fig. 8(a)), the color transfer results are shown in Figs. 8(b1-d1). Compared with the ground truth, the RMSEs are not a certain value and are different from part to part, which means this kind of color transfer is not stable. The average RMSE of these three tiles is 6.3%, which is a bit worse than our CFPM scheme for the same sample with the RMSE of 5.37%. And also it appears the chromatic aberration problem (Figs. 8(b1-d1)). Therefore, the color texture information of a LR full-FOV full-color donor is enough for the color transfer to the HR grayscale acceptor image.

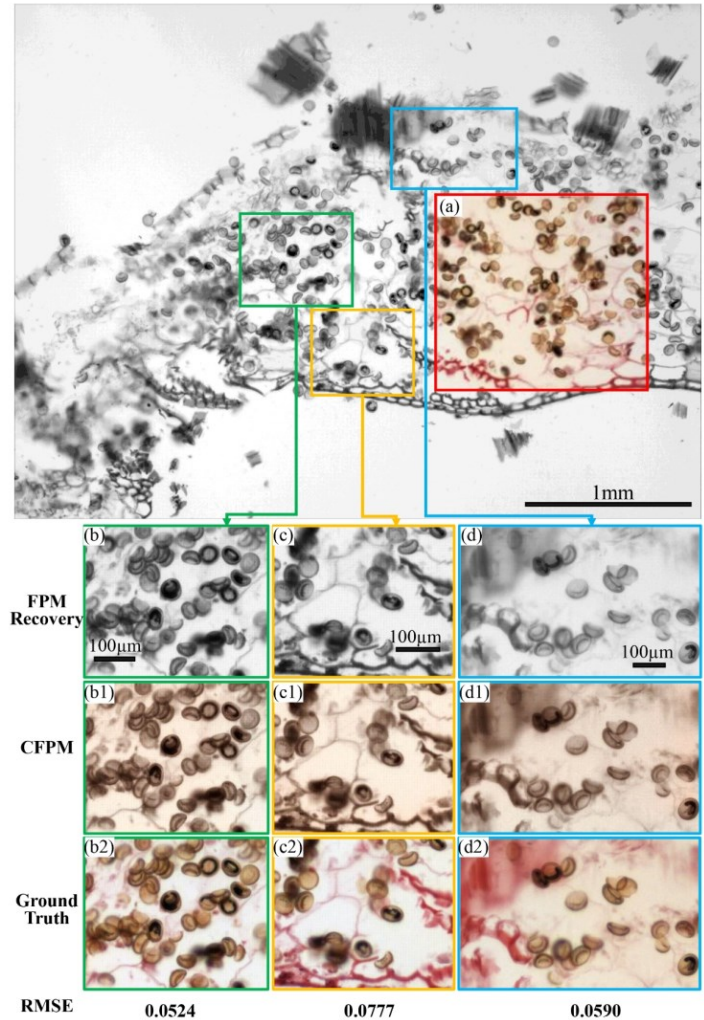

**Fig. 8** comparison of different color transfer scheme. (a) A HR donor image captured by a higher NA and magnification objective, therefore its FOV is much smaller. (b-d) Three random tiles as the acceptor images captured by the green channels, respectively. (b1-d1) Corresponding results of this kind of color transfer method, respectively. (b2-d2) Corresponding ground truth, respectively.

## 6   Conclusions

In summary, high-throughput full-color imaging is significant in digital pathology and FPM is a promising high-throughput computational imaging technique. Currently, the conventional full-color digital pathology based on FPM is still time-consuming due to the repeated experiments with tri-wavelengths. To solve this problem, four kinds of approaches are discussed and compared, and they may not really save recovery time or obtain enough accuracy. In this paper, we reported A color transfer FPM approach, termed CFPM. The color texture information of a LR full-color pathologic image is directly transferred to the HR grayscale FPM image captured by only a single wavelength, which may be regarded as a kind of unsupervised transfer learning. The color space of our FPM system based on the standard CIE-XYZ color model and display based on the standard RGB (sRGB) color space were established and illustrated in detail. The accuracy and efficiency of different FPM colorization schemes were compared with thirty different biological samples. The average RMSE of the conventional method and our CFPM is 5.3% and 5.7%, respectively, which means that the acquisition time is significantly reduced by 2/3 with the sacrifice of precision of only 0.4%. Also, the difficulty of CFPM to distinguish multiple dyes is also pointed out, which may be our future work. The demo code is released on our website for uncommercial use.

*Disclosures*

The authors have no relevant financial interests in this article and no potential conflicts of interest to disclose.

*Acknowledgments*

The authors acknowledge the National Natural Science Foundation of China (NSFC) (81427802 and 61377008).

**Yuting Gao** received her BE degree from Xi'an University of Architecture and Technology in 2019. Currently, she is a second-year PhD candidate at Xi'an Institute of Optics and Precision Mechanics (XIOPM), Chinese Academy of Sciences (CAS), China. Her research focuses on computational imaging. She is a student member of both OSA and SPIE.

**An Pan** received his BE degree from Nanjing University of Science and Technology (NJUST), China in 2014. Pan was a visiting graduate at Bar-Ilan University, Israel in 2016 and California Institute of Technology (Caltech), USA from 2018 to 2019. And he obtained his PhD in optical Engineering at Xi'an Institute of Optics and Precision Mechanics (XIOPM), Chinese Academy of Sciences (CAS), China in 2020. Currently, he is the head and a principal investigator (PI) of Pioneering Interdiscipline Center (PIC), State Key Laboratory of Transient Optics and Photonics (www.piclaboratory.com). His research focuses on computational imaging and space physics. He is the member of OSA, SPIE, and APS. He was awarded the 2019 OSA Boris P. Stoicheff

**Caption List**

**Fig. 1** The importance of true colorization of grayscale images. (a) Grayscale image of a cell section. (b) Pseudo-color image (c) True-color image of the cell section.

**Fig. 2** Inspiration of color matching and the principle of CFPM approach. (a) Match color: transfer the hue of donor image to the acceptor image. (b) Color transfer: transfer the color texture of donor image to the acceptor image.

**Fig. 3** Schematic diagram of CFPM. (a) Donor image: A LR full-color image with the same FOV of FPM captured by the same objective. (b) Acceptor: FPM recovery with a single wavelength. (c) CFPM final recovery.

**Fig. 4** CFPM setup with the illumination of LED board. (a) Schematic and its experimental photograph (b). (a1) A 32×32 programmable R/G/B LED matrix. (a2) The enlargement of a compact inverted microscope with light path diagram. It can be simply modified to realize multimodal imaging (a3). MO: microscope objective; TL: tube lens; M1 and M2: mirrors; BS: beam splitter; DM1 and DM2: dichroic mirror. (c) Standard color gamut chart of 1931 CIE-XYZ and our color space of FPM (blue triangle). (c1-c4) Wavelength and intensity calibration. He-Ne laser: 5.6 nm deviation tested by the spectrometer of Ocean Optics. Red LED: 630.1 nm of center wavelength, 20.8 nm of FWHM. Green LED: 515.0 nm of center wavelength, 38.0 nm of FWHM. Blue LED: 462.6 nm of center wavelength, 34.6 nm of FWHM.

**Fig. 5** Stained resting sporangia. (a) The LR donor image with entire FOV. (b) The FPM recovery image under green channel. (a1) A tile of (a). (b1-b3) Close-ups of FPM recovery images under red, green, blue channel, respectively. (c) Ground truth. (d1-d3) The corresponding results of CFPM respectively.

**Fig. 6** Stained lily bud cells. (a) The LR donor image with entire FOV. (b) The FPM recovery image under green channel. (a1) A tile of (a). (b1-b3) Close-ups of FPM recovery images under red, green, blue channel, respectively. (c) Ground truth. (d1-d3) The corresponding results of CFPM respectively.

**Fig. 7** Comparisons between three kinds of FPM colorization results. (a) RMSE curves of conventional R+G+B, our CFPM, and WMFPM with 30 pathologic or biological samples. (b) Ground truth of the 24th biological sample. (c-e) Recovery of corresponding CFPM, conventional R+G+B, and WMFPM respectively.

**Fig. 8** comparison of different color transfer scheme. (a) A HR donor image captured by a higher NA and magnification objective, therefore its FOV is much smaller. (b-d) Three random tiles as the acceptor images captured by the green channels, respectively. (b1-d1) Corresponding results of this kind of color transfer method, respectively. (b2-d2) Corresponding ground truth, respectively.